\documentclass[prb,aps,twocolumn,showpacs]{revtex4}

\usepackage{epsfig}

\begin{document}

\title{Site-dependent NMR Spin-lattice Relaxation in the Superconducting State of an Iron Pnictide Superconductor}

\author{L. Ma ${^1}$}
\author{J. Zhang $^{2}$}
\author{G. F. Chen $^{1} $}
\author{T.-L. Xia ${^1}$}
\author{J. B. He${^1}$}
\author{D. M. Wang${^1}$}
\author{W. Yu $^{1}$}
\email{wqyu_phy@ruc.edu.cn}

\affiliation{
$^1$Department of Physics, Renmin University of China, Beijing 100872, China\\
$^2$School of Energy and Power Engineering, North China Electric Power University, Beijing 102206, China\\
}
\date{\today}

\begin{abstract}

In a conventional superconductor, the spin-lattice relaxation rate on all nuclei should have the same temperature dependence below T$_C$. We performed $^{23}$Na, 
$^{75}$As, and $^{59}$Co NMR studies on single crystals of NaFe$_{0.95}$Co$_{0.05}$As, and found that spin-lattice relaxation rates show very different temperature 
dependent power-law behavior on three sites. We propose that such site-dependent behavior is due to the facts that the superconductor has two gaps of very different size.
The power-law exponent of each nucleus is affected by the strength of the hyperfine coupling to the small gap. We also found that the large superconducting gap on the 
cobalt site is smaller than on other two sites. It suggests a local suppression of the superconducting gap on the dopant site. 
\end{abstract}

\maketitle

The Hebel-Slichter formulation is very successful in analyzing the NMR spin-lattice relaxation rate (SLRR) of superconductors \cite{Hebel_PR_113_1504-1519, 
Hebel_PR_116_79}. Below $T_C$, the scheme can be used to determine the symmetry and the amplitude of the superconducting gap. For example, the low-temperature SLRR 
usually shows a gaped behavior with a coherence peak in a $s$-wave superconductor, and a $T^3$ power law behavior in a $d$-wave superconductor. In the newly discovered 
high-$T_C$ iron pnictide superconductors\cite{Hosono_Jacs_130_3296,Chen_Nature_453_761,Chen_PRL_100_247002,Ren_MRI_12_105}, which are probably multiple-gap
superconductors \cite{Mazin_PRL_101_057003,Cvetkovic_EPL_85_37002,seo_PRL_101_206404,Kuroki_PRL_101_087004,Wang_PRL_102_047005,Nakayama_EPL_85_67002,
Ding_EPL_83_47001}, however, the temperature dependence of the $^{75}$As SLRR below T$_c$ varies dramatically with dopings.  The SLRR of $^{75}$As follows a power-law 
temperature dependence by $1/T_1\sim T^n$, with the value of $n$ varies from  6 to 1.5 
\cite{Lee_PRB_80_144512,Grafe_PRL_101_047003,Kobayashi_JPSJ_78_073704,MuKuda_JPSJ_77_093704,Terasaki_JPSJ_78_013701,Fukazawa_JPSJ_78_033704,
Matano_EPL_83_57001,Kawasaki_PRB_78_220506,Matano_EPL_87_27012,Hammerath_CM_09123681,Nakai_CM_09080625,Yu_PRB_81_012503}. It is unclear if the different value of $n$ is 
caused by a change of the pairing symmetry with doping, a disorder scattering effect in an $S\pm$ gap superconductor 
\cite{Bang_PRB_79_054529,seo_PRB_79_235207,Chubukov_PRB_80_140515}, or other unknown mechanism. 

So far, NMR studies in iron pnictides were mostly performed on $^{75}$As. From the Hebel-Slichter formulation, the NMR SLRR of all nuclei on the same material should have 
the same temperature dependence, regardless of s-wave or d-wave superconductivity. The underlying physics is that the SLRR is primarily determined by electron 
excitations, with the electron density of states (DOS) depending on the gap symmetry and the gap amplitude. In this letter, we present our NMR studies on high-quality 
NaFe$_{0.95}$Co$_{0.05}$As superconductors. The NMR is performed on three nuclei, $^{23}$Na, $^{75}$As and $^{59}$Co. Below $T_C$,  the SLRR show a power-law like 
behavior on all sites, but with very different power-law exponent $n$ on three nuclei. Our analysis indicates that such non-scale behavior is well understood by two 
superconducting gaps but with different gap sizes. We further found that the large gap on the cobalt site is much smaller compared with other sites. Since cobalt serves 
as a dopant, such spatial variation of the gap value suggests a local suppression of gap amplitude on the dopant site.

\begin{figure}
\includegraphics[width=8cm, height=6cm]{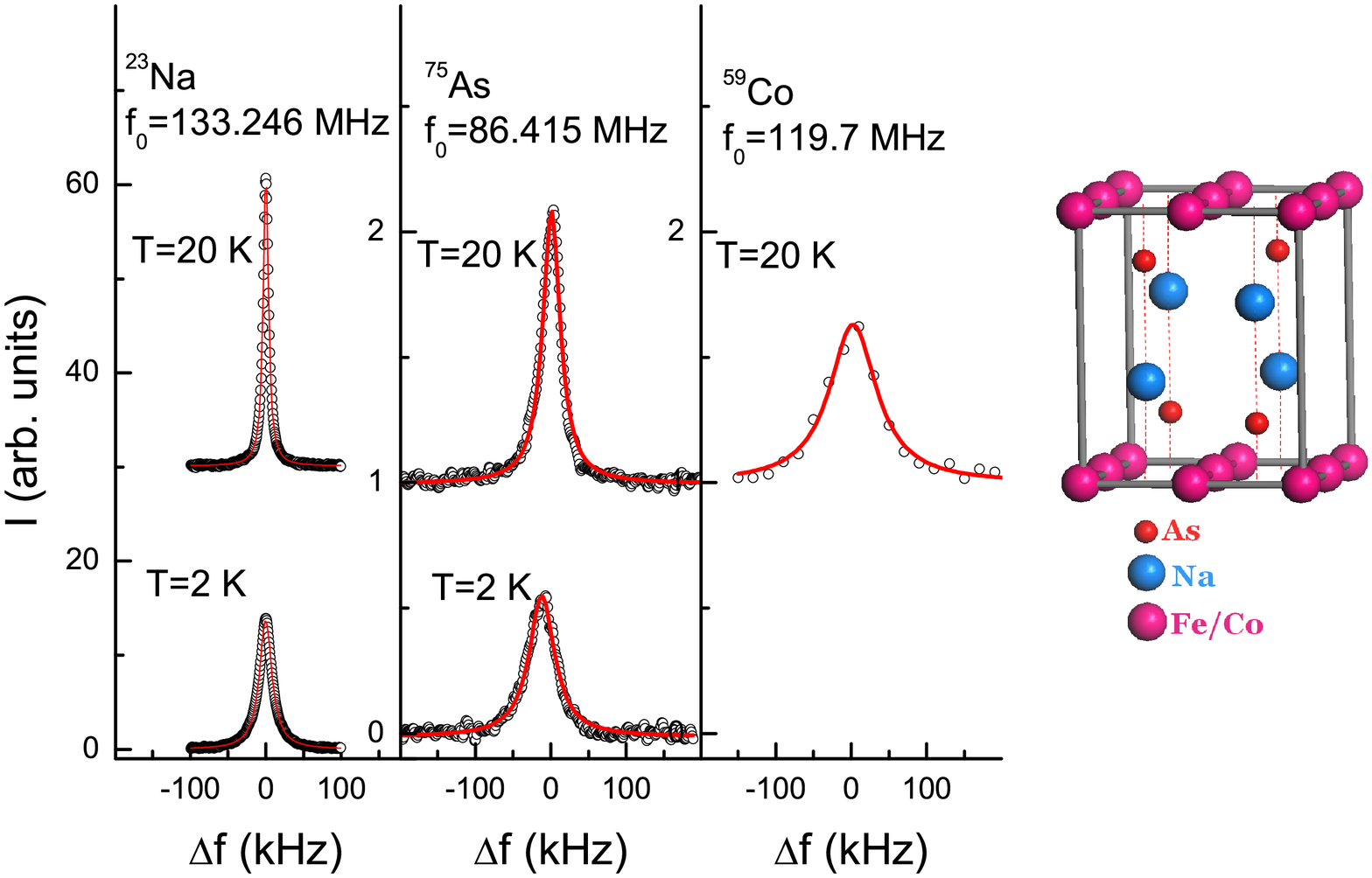}
\caption{\label{spectrum}(color online) Left panel: The NMR spectrum (dotted lines) of $^{23}$Na, $^{75}$As, and $^{59}$Co of a NaFe$_{0.95}$Co$_{0.05}$As single crystal 
at $T=$20 K (above $T_C$) and $T=$2 K (below $T_C$) with an applied field of 11.85 Tesla along the c-axis. The solid lines are the Lorentz function fitting. Right panel: 
A sketch of the NaFe$_{1-x}$Co$_x$As unit cell.}
\end{figure}

The  growth of our electron-doped NaFe$_{0.95}$Co$_{0.05}$As superconducting single crystals has been reported previously \cite{Chen_CM_10013311}. The same sample is used 
here, and the superconducting transition is indicated by the sharp transition from the magnetization and resistivity at $T_C\approx 18$ K\cite{Chen_CM_10013311}. Our NMR 
measurements were performed on $^{23}$Na, $^{75}$As, and $^{59}$Co, with 6 Tesla and 11.85 Tesla magnetic field, and temperatures down to 1.5 K. 

As shown in Fig.~\ref{spectrum}, Na, Co and As atoms reside on three typical positions in the lattice \cite{LiSL_NaFeAs}. Their typical NMR spectra are also shown in 
Fig.~\ref{spectrum}, and all lineshapes are fit by Lorentz function. At $T=$20 K, a temperature slightly above $T_C$, the full width of half maximum (FWHM) of the 
spectrum is about 8 kHz, 20 kHz and 34 kHz for $^{23}$Na, $^{75}$As and $^{59}$Co, respectively. The narrow linewidth at such a low temperature and a high field indicates 
that the quality of our sample is very high. Below $T_C$, the spectrum is significantly broadened due to the distribution of magnetic fields around the vortex core 
\cite{Brandt_PRB_37_2349}.


Our spin-lattice relaxation is measured by an inversion pulse method. The SLRR in a solid follows 
\begin{eqnarray}
 1/T_{1}T= \frac{\pi k_B \gamma^2_n}{(\gamma _e \hbar)^2}  {\sum _q}A^2_{hf}(q) \frac{\chi^{''} _{\perp}(q,\omega)}{\omega},  \label{a1}
\end{eqnarray}
where $A_{hf}(q)$ is the hyperfine coupling constant, and $\chi^{''} _{\perp}(q,\omega)$ is the imaginary part of the electronic dynamical susceptibility perpendicular to 
the magnetic field. 

\begin{figure}
\includegraphics[width=9cm, height=7cm]{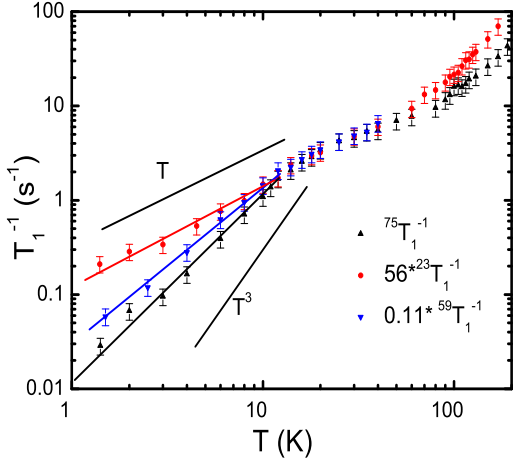}
\caption{\label{invt1s1}(color online) The temperature dependent spin-lattice relaxation of $^{23}$Na, $^{75}$As and $^{59}$Co of a NaFe$_{0.95}$Co$_{0.05}$As single 
crystal under a 11.85 Tesla magnetic field along the c-axis.  } 
\end{figure}

The SLRR of a NaFe$_{0.95}$Co$_{0.05}$As single crystal under a 11.85 Tesla magnetic field is shown in Fig.~\ref{invt1s1}. The 6 Tesla data (not shown in figure) gives 
identical results, showing no field-dependent behavior. The SLRR results show that the hyperfine coupling orders as  $^{59}A_{hf}>^{75}A_{hf}>^{23}A_{hf}$. For comparison 
purpose, $1/^{23}T_1$ and $1/^{59}T_1$ are multiplied by a factor of 56 and 0.11, respectively. Above $T_C$, the SLRRs of three sites show similar temperature dependence. 
From 80 K to 200 K, $1/T_1T$ increases rapidly with temperature, and such behavior is known as a pseudogap-like phenomena reported in other iron 
pnictides\cite{Ning_PRL_104}. Between 80 K and $T_c$, the SLRR of all sites is weakly temperature dependent, which is an indication of spin 
fluctuations\cite{Ning_PRL_104}.   
  
The Superconducting transition is shown by a drop of $1/T_1$ on all three sites. Similar to reports on other iron pnictides, no coherence peak is found in 
NaFe$_{0.95}$Co$_{0.05}$As, suggesting an unconventional superconductor  
\cite{Lee_PRB_80_144512,Grafe_PRL_101_047003,Kobayashi_JPSJ_78_073704,MuKuda_JPSJ_77_093704,Terasaki_JPSJ_78_013701,Fukazawa_JPSJ_78_033704,
Matano_EPL_83_57001,Kawasaki_PRB_78_220506,Matano_EPL_87_27012,Hammerath_CM_09123681,Nakai_CM_09080625,Yu_PRB_81_012503}. Below $T_C$, $1/^{23}T_1$, $1/^{59}T_1$ and 
$1/^{75}T_1$ show power-law-like temperature dependence ($1/T_{1}\approx T^n$), but the power law exponent are different with $n\approx 2$ for $^{75}$As, $n\approx 1.7$ 
for $^{59}$Co, and $n\approx  1.0$ for $^{23}$Na. The sequential decrease of $n$ suggests more contributions from the low energy excitations.   
 
These SLRRs data cannot be fit by a single $s$-wave or a single $d$-wave function. Here we attempt to fit the data by a two-gap function, like in other iron arsenide 
superconductors \cite{Matano_EPL_83_57001}. The spin-lattice relaxation for a two-gap Fermi liquid superconductor is approximated by,
\begin{eqnarray}
1/T_{1n}=2 \displaystyle{\sum _{i=1,2}} {\int^\infty_0 (A^i_{hf} N^i_{0}(E))^2f(E)(1-f(E))\, dE}  &&\label{a2} \\
T_{1n}/T_{1s}=\frac{2}{k_BT}\displaystyle{\sum _{i=1,2}} {\int^\infty_0 (d^{i}_{S}(E))^2f(E)(1-f(E))\, dE}  &&\label{a3}
\end{eqnarray}
where $T_{1n}$ and $T_{1s}$ are the spin-lattice relaxation time of the normal state and the superconducting state, respectively. $N^i_{0}$ (i=1,2) is the normal state 
electron density of states (DOS) of each band, $A^i_{hf}$ is the hyperfine coupling of the nucleus to each band, and $f(E)$ is the Fermi-Dirac distribution function with 
$f(E)=1/(e^{-E/k_BT}+1)$. Assuming a hyperbolic Fermi surface with $N^1_{0}$ and $N^2_{0}$ as constants, we define an {\it effective} normal sate DOS $d^i_0\equiv 
A^i_{hf} N^i_{0}/D_0 \equiv A^i_{hf} N^i_{0}/(A^1_{hf} N^1_{0}+ A^2_{hf} N^2_{0})$ with $d^1_0+d^2_0=1$, and an {\it effective} superconducting state DOS $d^i_S(E)\equiv 
A^i_{hf} N^i_{S}(E)/D_0$ with  $N^i_{S}$ as the DOS in each superconducting band. Then $d^i_S$ is simplified as $d^i_S\equiv d^i_{0} N^i_{S}/ N^i_{0}$, with $N^i_{S}/ 
N^i_{0}$ as a functions of two superconducting gaps $\Delta^1_S$ and $\Delta^2_S$, depending on the detailed gap symmetry. In the end, Eq.~\ref{a2} and Eq.~\ref{a3} 
contain only four fitting parameters, $D_0$, $d^1_{0}$, $\Delta^1_S$ and $\Delta^2_S$ .

Since the system has spin fluctuations, we first fit the normal state SLRR data by $1/T_1T\sim A/(T+\Theta)$ between $T_C$ and 80 K.  The $1/T_1$ data at all temperatures 
are then multiplied by $(T+\Theta)/\Theta$ to remove the spin fluctuation effect so that the Fermi liquid assumption in Eq. \ref{a2} holds. Finally, we fit the data by 
Eq.~\ref{a2} (above $T_C$) and Eq.~\ref{a3} (below $T_C$) with the same parameters.

\begin{figure}
\includegraphics[width=9cm, height=10cm]{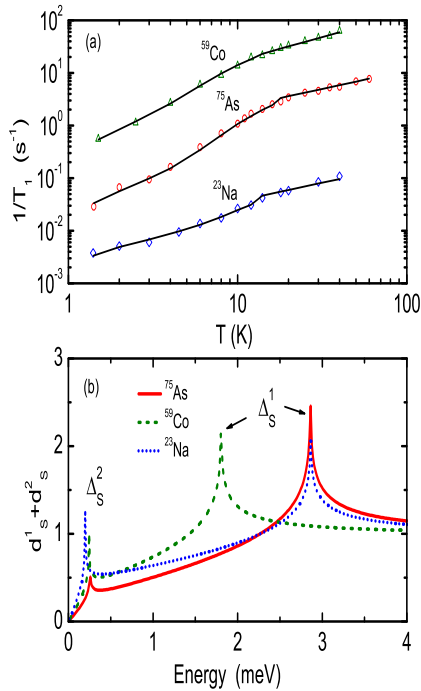}
\caption{\label{t1fit}(color online) (a) The low-temperature $^{23}$Na, $^{75}$As, and $^{59}$Co spin-lattice relaxation  rate (symbols) and the fittings (solid lines)  
by the two-gap $d$-wave symmetry as described in the text. (b) The effective electron density of states on $^{23}$Na, $^{75}$As, and $^{59}$Co sites obtained from the 
fittings. }
\end{figure}


\begin{table}[h]
\caption{\label{t1}The two-gap $d$-wave fitting parameters of the spin-lattice relaxation rate of $^{23}$Na, $^{75}$As, and $^{59}$Co.}
\begin{tabular}{p{1cm}p{1cm}p{2.5cm}p{2.5cm}}
\hline
	site	& $d^1_{0}$  &  $\Delta^1 _{S}$ (meV) &  $\Delta^2 _{S}$ (meV) \\
\hline
$^{75}$As	& 	0.78	&  2.84$\pm$0.15		 & 	0.26$\pm$0.05\\ 
\hline
$^{59}$Co	&  0.68		&  1.81$\pm$0.15		 & 	0.24$\pm$0.05\\
\hline
$^{23}$Na	&  0.57		&  2.84$\pm$0.15		 & 	0.20$\pm$0.05\\	
\hline	
\end{tabular}
\end{table}

We first use a two-gap $d$-wave fitting, assuming the gap varies with angle by $\Delta _i( \phi )=\Delta^i _{S} cos(2\phi)$. The effective DOS, by averaging out the angle 
dependence, is given by $d^i_{S} =  d^i_{0}\frac{2}{\pi}K(\frac{\Delta^{i2}_{S}}{E^2})$ for $E>\Delta^i _{S}$,  and  $d^i_{S} = d^i_{0} \frac{2}{\pi} \frac{E}{\Delta^i 
_{S}} K (\frac{E^2}{\Delta^{i2}_{S}})$ for $E<\Delta^i _{S}$, where $K(x)$ is an elliptic function. A good fitting is obtained for the three nuclei as shown in 
Fig.~\ref{t1fit}(a), and the total effective DOS ($=d^1_{S}+d^2_{S}$) below $T_C$ are depicted in Fig.~\ref{t1fit}(b). The fitting parameters for each nucleus are listed 
in Table.~\ref{t1}. The close values of the large gap $\Delta^1 _{S}$ and the small gap $\Delta^2 _{S}$ among all nuclei suggest that our fittings are physically 
reasonable.

As shown in Fig.~\ref{t1fit}(b), the effective DOS is peaked at two energies corresponding to the gap value $E=\Delta^1 _{S}$ and $E=\Delta^2 _{S}$. The small gaps are 
similar for all sites with $2\Delta^2 _{S}/k_BT_C\approx 0.3$. The large gap $\Delta^1_{S}$ are similar on the As and the Na sites with $2\Delta^1 _{S}/k_BT_C\approx 
3.7$, but is much smaller on the Co site with $2\Delta^1_{S}/k_BT_C\approx 2.3$. The smaller $\Delta^1 _{S}$ on the cobalt site suggests a spatial suppression of gap 
close to it. Since Co serves as a dopant in the lattice, the gap suppression is probably caused by a local scattering effect from doping. Similarly, a local gap 
suppression by dopant has been reported by STM studies in the cuprates \cite{Hirschfeld_STM}. 

In Table \ref{t1}, a systematic decrease of $d^1_{0}$ ($E=\Delta^1 _{S}$) is seen in the order of $^{75}$As, $^{59}$Co, and $^{23}$Na, which is reversely demonstrated by 
the increase of the effective DOS close to small gap position ($E=\Delta^2 _{S}$) in Fig.~\ref{t1fit}(b). Physically, it suggests that the $^{23}$Na senses more low 
energy excitations than other sites, which explains the smaller power law exponent $n$ of the SLRR on $^{23}$Na. With $d^1_0$ rewritten as $d^1_0\equiv 
1/(1+(A^2_{hf}/A^1_{hf})(N^2_{0}/N^1_{0}))$, $d^1_0$ is determined by the ratios of the actual DOS in two bands $N^1_0/N^2_0$ and the ratio of the hyperfine coupling to 
two bands $A^1_{hf}/A^2_{hf}$. It is reasonable to assume that  $N^1_0/N^2_0$ does not vary with nucleus sites, and is determined by the two electron bands. 
$A^1_{hf}/A^2_{hf}$, however, could vary with nucleus site, since the hyperfine coupling is momentum dependent (see Eq.~\ref{a1}), and the Fermi surfaces of different 
bands of the iron arsenides are separated in the Brillouin zone \cite{Singh_PRB_78_094511}. Then the final conclusion is reached that the smaller power-law exponent $n$ 
of the SLRR is caused by a stronger local hyperfine coupling of the nucleus to the band with the small gap. 

\begin{figure}
\includegraphics[width=8cm, height=5cm]{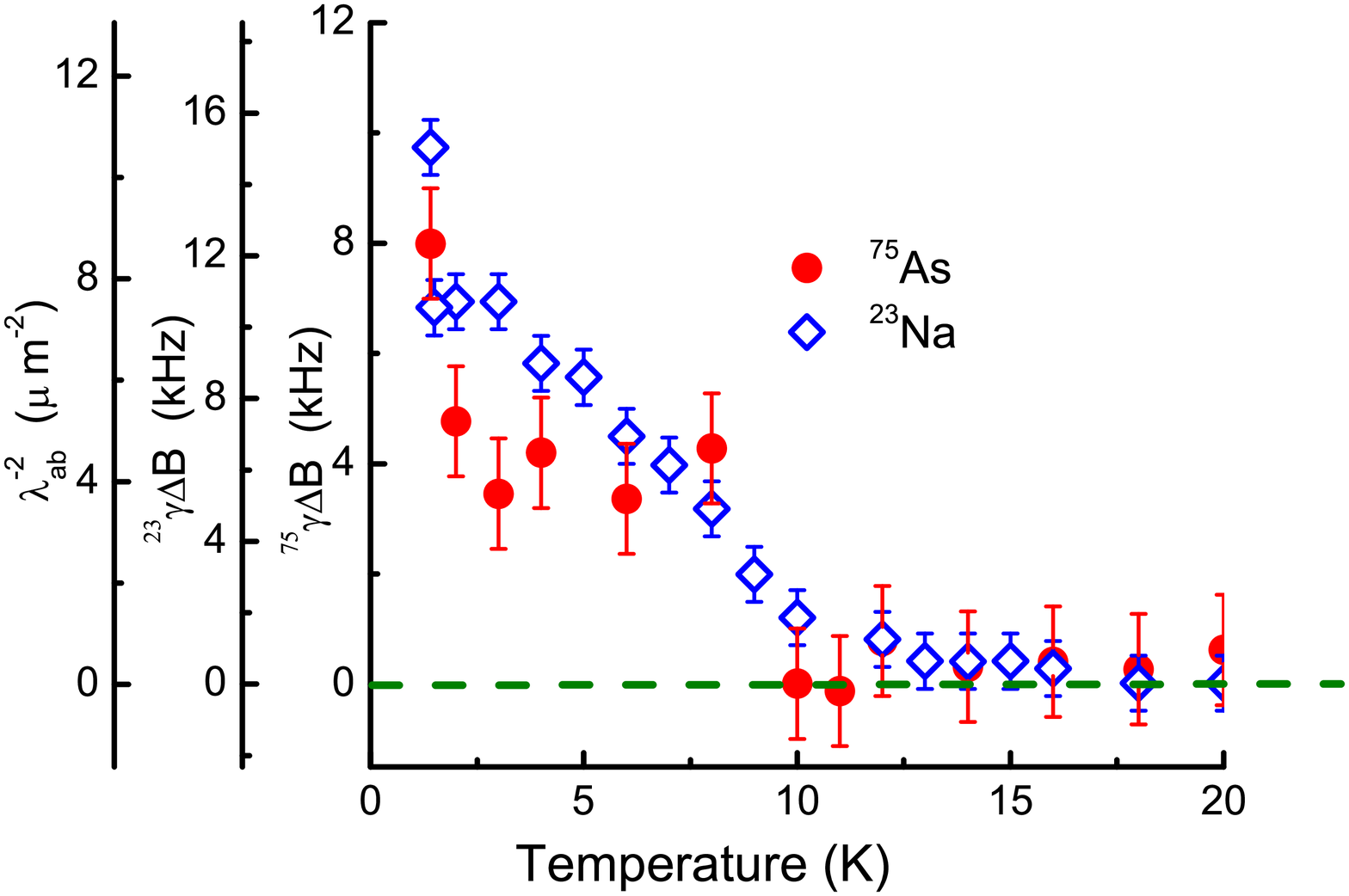}
\caption{\label{rous}(color online) The increase of the NMR linewidth of $^{23}$Na and $^{75}$As spectrum below $T_C$. Both linewidth are scaled to $\lambda^{-2}_{ab}$ as 
shown by the vertical axis (see text), where $\lambda _{ab}$ is the penetration depth.}
\end{figure}

Before going further, we confirm that our SLRR for all nuclei are intrinsic for several reasons. First, the spin recovery for each nucleus has only one $T_1$ component 
below $T_C$. Second, the normal state $1/T_1$ on all sites shows a similar temperature dependence, which is consistent with other electron-doped iron arsenide 
\cite{Ning_PRL_104}. Finally and most importantly, we confirmed that $^{23}$Na and $^{59}$Co signals are excited from the same region around the vortex core within the 
penetration depth. The penetration depth is evaluated from the spectrum broadening below $T_C$ due to the local distribution of fields around a vortex core 
\cite{Brandt_PRB_37_2349}. For each nucleus, the penetration depth $\lambda _{ab}$ is calculated by $ \Delta B= \Delta f/\gamma _n \approx 0.0609\phi _0/\lambda^2_{ab}$ 
below T$_C$ \cite{Brandt_PRB_37_2349}, where $\phi _0$ is the quantum flux, $\Delta f$ is the linwwidth, and $\gamma _n$ is the gyromagnetic factor of the nucleus.  As 
shown in Fig.~\ref{rous}, the penetration depth approaches $\lambda _{ab}\approx$0.3 $\mu$m as $T\rightarrow $0 K from both $^{23}$Na and $^{75}$As measurements. Such 
penetration depth is typical for iron arsenide with a similar $T_c$\cite{Carlo_musr}.

Therefore, our analysis suggests that the temperature dependence of the NMR SLRR below T$_C$ is affected by two effects. First, the gap is locally suppressed by the 
disorder effect on the dopant site. Second, the different power-law exponent $n$ of the SLRR on each nucleus is well understood in a superconductor with a large gap and a 
small gap. It is worthwhile to mention that ARPES may not be able to resolve the small gap. Although our fitting scheme is based on a d-wave symmetry, the conclusion is 
also valid on a two-gap superconductor with other symmetries. For example, we believe an anisotropic two-gap $s$-wave fitting should also work, if a similar electron DOS 
is obtained as that of the $d$-wave fitting. The proposed $s_\pm$ gap symmetry with strong scatterings \cite{Bang_PRB_79_054529,Chubukov_PRB_78_134512,seo_PRB_79_235207} 
may also work. 

Our conclusion that the power-law behavior of the SLRR varies with local hyperfine couplings and local gap amplitude may shed light on $^{57}$As NMR studies on other iron 
pnictides in two aspects. First, the local gap suppression from the dopant should be stronger with the increase of the doping. As a result, doping may cause more low 
energy excitations, and lower the power-law exponent as shown on the cobalt site. Second, with the increase of doping, the local hyperfine couplings $A_{hf}(q)$ of 
$^{57}$As on the Fermi surface may also vary with doping, because the Fermi surface changes dramatically with doping in iron arsenides \cite{Sato_PRL_103_047002}. In both 
cases, the change of the power-law exponent $n$ are not necessarily related to the change of the gap symmetry.

In summary, our NMR study on NaFe$_{0.95}$Co$_{0.05}$As single crystals shows that the SLRR on $^{23}$Na, $^{59}$Co, and $^{75}$As sites has very different temperature 
dependent power-law behavior below $T_C$. Our analysis suggests that such non-scale behavior is well described by the Hebel-Slichter formulation based on a two-gap 
superconductor with a large gap and a very small gap, regardless of the detailed gap symmetry. We also found that the large gap on the Co-dopant site is strongly 
suppressed, probably due to the disorder effect. To our knowledge, this is the first report of a local gap suppression on the dopant site, and the effect may be verified 
by scanning probes such as STM. 

The Authors acknowledge Dr. Wei Bao, Qiang Han, Zhongyi Lu, Tiesong Zhao for helpful discussions. W.Y. and G.F.C are supported by the National Basic Research Program of 
China. W.Y. is also supported by Program for New Century Excellent Talents in University.


\end{document}